\documentstyle[psfig]{mn0}

\begin{document}

\parskip 3pt
\topmargin 5pt

\title[Peaks of afterglow light-curves]
      {Peaks of optical and X-ray afterglow light-curves}

\author[Panaitescu, Vestrand \& Wo\'zniak]{A. Panaitescu, W.T. Vestrand, P. Wo\'zniak \\
       Space \& Remote Sensing, MS B244, Los Alamos National Laboratory,
       Los Alamos, NM 87545, USA}

\maketitle

\begin{abstract}
\begin{small}
 The peaks of 30 optical afterglows and 14 X-ray light-curves display a good anticorrelation of the 
 peak flux with the peak epoch: $F_p \propto t_p^{-2.0}$ in the optical, $F_p \propto t_p^{-1.6}$ in 
 the X-ray, the distributions of the peak epochs being consistent with each other.
 We investigate the ability of two forward-shock models for afterglow light-curve peaks -- an observer 
 location outside the initial jet aperture and the onset of the forward-shock deceleration -- to account 
 for those peak correlations. 
 For both models, the slope of the $F_p - t_p$ relation depends only on the slope of the afterglow spectrum. 
 We find that only a conical jet seen off-aperture and interacting with a wind-like medium can account 
 for both the X-ray peak relation, given the average X-ray spectral slope $\beta_x = 1.0$, and for the
 larger slope of the optical peak relation.
 However, any conclusion about the origin of the peak flux -- peak epoch correlation is, at best, tentative,
 because the current sample of X-ray peaks is too small to allow a reliable measurement of the $F_p - t_p$ 
 relation slope and because more than one mechanism and/or one afterglow parameter may be driving that 
 correlation.
\end{small}
\end{abstract}


\section{Introduction}

 The power-law fall-off of the GRB afterglow flux is the most-often seen feature and was predicted by 
M\'esz\'aros \& Rees (1997). Long-monitored afterglow light-curves also display at least one break.
Early (1-10 ks from trigger) breaks in the X-ray light-curve have been interpreted as due to energy
being added to the afterglow blast-wave (Nousek et al 2006). Later (0.3-3 d) breaks in the optical 
light-curve have been predicted by Rhoads (1999) in the framework of a tightly collimated afterglow outflow. 
 
 A less-often encountered feature of afterglows is the peak displayed by some optical light-curves
at early times (up to 1 h after trigger). A case by case modelling of those peaks could provide a test
of the possible models for a light-curve peak, as was done, for instance, for the jet model using 
the optical light-curve breaks seen at later times. We have investigated models for light-curve peaks
only in a general sense, by assessing whether a given model can explain the strong anticorrelation 
observed between the peak flux $F_p$ and the peak epoch $t_p$. Using {\sl numerical} calculations of the
reverse and forward-shock synchrotron light-curves for jets seen at various angles, we (Panaitescu \& 
Vestrand 2008) have found that that model can account qualitatively for the $F_p \propto t^{-2.7}$ 
peak anticorrelation measured for a dozen afterglows with optical light-curve peaks. {\sl Analytical} 
results for the peak flux and epoch expected at the onset of the forward-shock deceleration have been 
used by us (Panaitescu \& Vestrand 2011) to test quantitatively the $F_p - t_p$ relation for optical peaks.

 In this work, we continue to investigate the peak models mentioned above, by applying {\sl analytical} 
results for the $F_p - t_p$ relation expected for the synchrotron emission from the {\sl forward-shock}
to afterglows with {\sl X-ray peaks}. The fraction of Swift X-ray afterglows displaying a peak is much 
smaller than for optical afterglows because the X-ray peak is most often missed, being overshined by the 
prompt GRB emission. We have found about a dozen of X-ray light-curves with peaks among the several 
hundred X-ray afterglow light-curves in the XRT database, most of which are accompanied by a spectral 
hardness evolution that indicates that the emerging X-ray light-curve peak has an origin (= afterglow)
different than the GRB prompt emission. The importance of X-ray peaks lies in that, unlike for optical 
peaks, the slope of the X-ray afterglow spectrum was measured and can be used for model testing, as all 
light-curve peak models lead to an $F_p \propto t_p^{-\gamma}$ relation with an exponent that depends 
on the spectral slope. As we shall see, the $F_p - t_p$ peak relation is steeper for optical peaks than 
for X-ray peaks, which we will use to further test the two light-curve peak models.

\section{Optical peaks}

 The sample of afterglows with optical light-curve peaks used in this work contains 19 afterglows
presented in Panaitescu \& Vestrand (2011), which are the afterglows with peaks observed until 2009  
and for which the peak $fwhm$ (full width at half maximum) is less than a decade (1 dex) in time. 
We add to that set 12 other peaks, some with $fwhm$ larger than 1 dex (they were previously considered 
plateaus or of uncertain type). For a few afterglows, the peak does not start at the first measurement
but appears after a flat or a decreasing optical flux; 110205 displays two clear peaks, both have 
included in the current sample.

 For that set of 31 peaks, the $z=2$ optical fluxes ($F_{2eV}$) were calculated assuming
an optical spectral slope $\beta_o = 0.75$. Accounting for extinction by dust in the host galaxy 
may lead to intrinsic optical fluxes that are a factor up to few brighter, however the $z=2$ peak 
fluxes of our sample span 5 decades (without 0606014), thus the error in the resulting relation 
between peak flux and peak epoch, owing to the (unaccounted for) host extinction, should be small.

\begin{figure*}
\centerline{\psfig{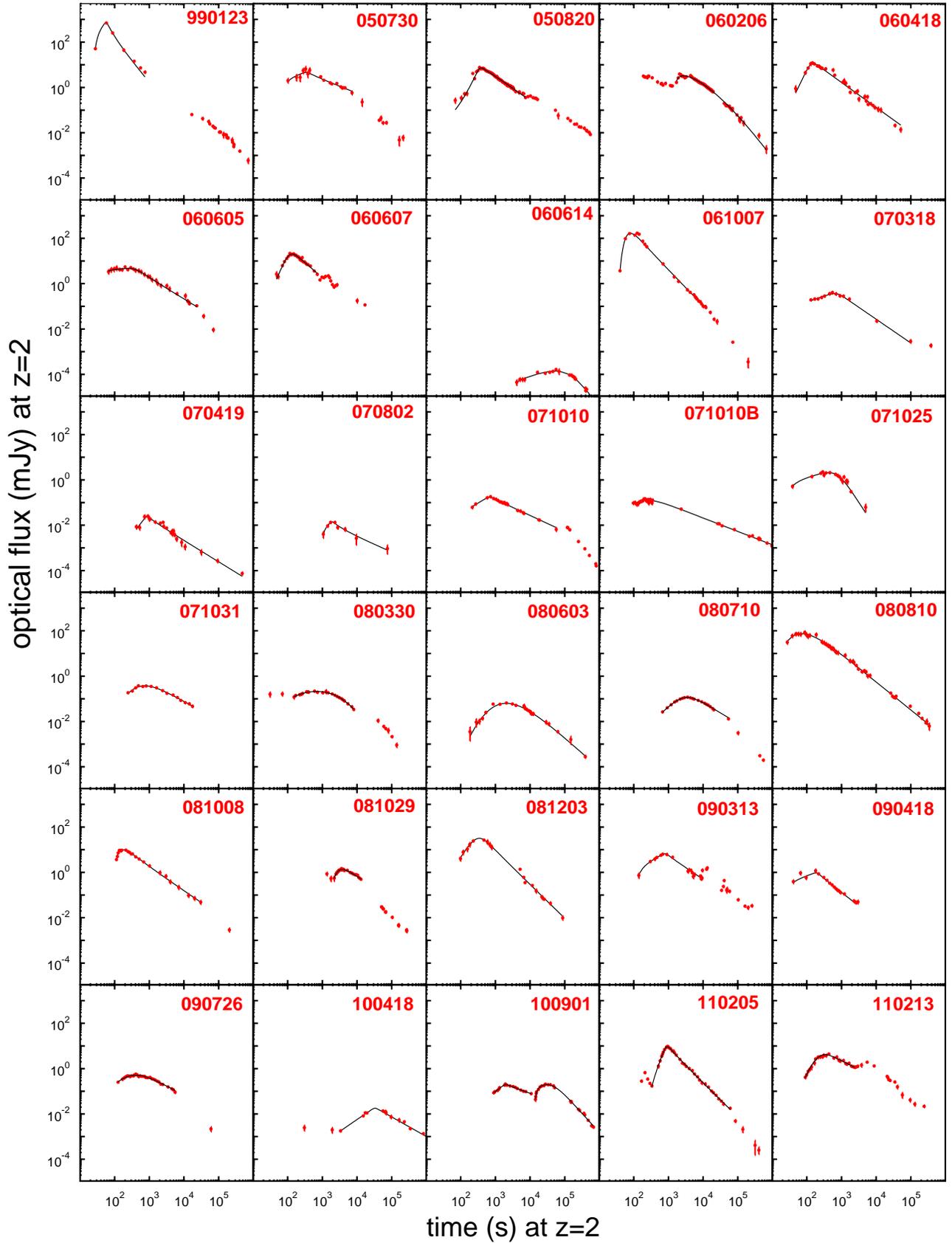}}
\caption{ Optical measurements for 30 afterglows and smoothly broken power-law fits to 31 peaks, 
   obtained with the function given in equation (\ref{fit}). Note that not all existing measurements
   have been fit, but only those that represent a light-curve peak. }
\label{allopt}
\end{figure*}

 To measure the epoch $t_p$ and flux $F_p$ at the light-curve peak, the optical light-curves
have been fit with a smoothed broken power-law
\begin{equation}
 F_\nu = f \left[ \left( \frac{t-t_o}{t_b} \right)^{-\xi\alpha_r} + 
                  \left( \frac{t-t_o}{t_b} \right)^{\xi\alpha_d} \right]^{-1/\xi}
\label{fit}
\end{equation}
with $\xi$ determining the sharpness of the transition between the asymptotic $F_\nu \propto t^{\alpha_r}$
rise and the $F_\nu \propto t^{-\alpha_d}$ decay (the larger $\xi$, the peakier is the light-curve), 
$t_b$ setting the epoch when the transition between these decays occurs (but $t_b$ is only a very 
rough estimate of the peak epoch $t_p$). We allow for a shift $t_o$ for the time when the 
optical afterglow emission begins relative to the GRB trigger (from when $t$ is measured), 
with $t_o > 0$ allowing for an optically emitting outflow released after the GRB-producing outflow,
 
 The data for the optical afterglows are shown in Figure \ref{allopt}, together with the broken power-law 
fits. The relevant best-fit parameters ($\alpha_r$, $\alpha_d$, $t_o$) to the optical peaks are given in 
Table \ref{otable}. 
The peak flux and epoch are not parameters of the fit, thus they were calculated from the best-fit 
parameters. With the exception of $t_o$, uncertainties were not determined, but we provide here an 
estimate of those uncertainties for the parameters of interest. 
 The $1\sigma$ error of $F_p$ and $t_p$ is less than $\epsilon = 0.15$. The exact value of 
$\epsilon(F_p)$ has almost no effect on the following $F_p - t_p$ relation, while the uncertainty 
on $t_p$ is much less than that of $t_o$, hence $\epsilon(t_o)$ determines the uncertainty in the 
true peak epoch ($t_p - t_o$).  

\begin{table*}
\caption{Parameters of the smoothed broken power-law best-fit (eq \ref{fit}) to $z=2$ optical 
       light-curves with peaks (Fig \ref{allopt}) and other quantities that characterize those peaks }
\begin{tabular}{lcccccccccccccccc}
   \hline \hline
 GRB     &$N$& $t_1$&$\alpha_r$&$t_p$ &$fwhm$& $F_p$ &$\alpha_d$&$t_{end}$&$\chi^2_\nu$&  Refs. \\
afterglow&   &  (s) &          &  (s) & (dex)& (mJy) &           &   (ks)   &                &        \\
         &(1)&  (2) &   (3)    &  (4) &  (5) &  (6)  &     (7)   &    (8)   &    (9)         &  (10)  \\
  \hline
 990123 &  7 &  28  & 0.8      & 56   & 0.4 & 700   & 1.63      & 0.75      & 3.09    &  Ga9,C9     \\
 050730 & 17 & 100  & 0.5      & 330  & 0.8 & 4.6   & 0.61      & 7.2       & 2.88    &  P6        \\
 050820 & 81 &  70  & 3.0      & 390  & 0.6 & 7.3   & 1.04      & 7.4       & 1.50    &  C6,V6     \\
 060206 &117 & 1700 & 0.3      & 2600 & 0.7 & 3.3   & 1.77      & 660       & 0.72    &  M6,W6,C7,S7   \\
 060418 & 28 &  48  & 2.6      & 150  & 0.6 & 12    & 1.12      & 14        & 1.85    &  Mo7       \\
 060605 & 29 &  67  & 0.1      & 260  & 1.0 & 4.7   & 0.92      & 24        & 1.04    &  F9        \\
 060607 & 28 &  47  & 3.6      & 130  & 0.6 & 20    & 1.34      & 0.71      & 1.53    &  Mo7,Z8,N9 \\
 060614 & 25 & 4100 & 0.8      & 49000& 1.5 &0.00014& 4.82      & 1300      & 1.40    &  D6,G6,Ma7 \\
 061007 & 20 &  40  & 1.6      & 82   & 0.5 & 170   & 1.53      & 9.90      & 0.55    &  Mu7,Y7    \\
 070318 & 13 & 130  & 0.5      & 640  & 1.2 & 0.37  & 1.05      & 99        & 0.88    &  Ch8      \\
 070419 & 21 & 420  & 1.4      & 850  & 0.5 & 0.024 & 0.94      & 480       & 1.77    &  M9       \\
 070802 &  8 & 1070 & 0.6      & 1900 & 0.5 & 0.014 & 0.68      & 75        & 1.68    &  K8       \\
 071010 & 21 &  220 & 0.6      & 690  & 0.9 & 0.18  & 0.71      & 57        & 0.31    &  Co8      \\
 071010B& 34 &  97  & 1.3      & 250  & 1.5 & 0.13  & 0.59      & 1140      & 0.83    &  W8       \\
 071025 & 17 &  39  & 0.4      & 420  & 1.2 & 2.1   & 2.36      & 5.0       & 1.69    &  P10      \\
 071031 & 13 & 240  & 0.2      & 840  & 1.0 & 0.36  & 0.80      & 18        & 1.02    &  K9a      \\
 080330 & 23 & 150  & 0.7      & 690  & 1.5 & 0.21  & 4.04      & 8.1       & 0.23    &  Gu9      \\
 080603 & 25 & 180  & 3.4      & 2100 & 1.3 & 0.063 & 1.37      & 390       & 1.04    &  G11      \\
 080710 & 21 & 670  & 1.6      & 3600 & 1.1 & 0.11  & 1.06      & 54        & 0.41    &  K9b      \\
 080810 & 42 &  27  & 0.8      & 64   & 1.0 & 75    & 1.23      & 340       & 1.27    &  P9       \\
 081008 & 23 & 110  & 0.7      & 180  & 0.6 & 9.9   & 1.04      & 30        & 2.31    &  Y10      \\
 081029 & 28 & 2200 & 1.8      & 3800 & 0.8 & 1.4   & 1.05      & 13        & 0.63    &  N11,H12  \\
 081203 & 17 &  99  & 3.5      & 330  & 0.7 & 29    & 1.61      & 87        & 0.44    &  K9       \\
 090313 & 19 & 140  & 0.7      & 780  & 0.7 & 6.6   & 1.01      & 9.3       & 1.10    &  M10      \\
 090418 & 17 &  41  & 0.3      & 190  & 0.7 & 0.92  & 1.16      & 2.7       & 0.86    &  H9       \\
 090726 & 32 & 130  & 0.8      & 390  & 1.2 & 0.50  & 1.10      & 5.5       & 0.71    &  S10      \\
 100418 & 14 & 3400 & 1.0      & 33000& 0.8 & 0.018 & 0.86      & 2800      & 2.19    &  M11      \\
100901(1)&16 & 1900 & 1.2      & 2000 & 1.0 & 0.20  & 0.58      & 11        & 0.51    &  V10      \\
100901(2)&23 & 14000& 0.4      & 30000& 0.8 & 0.20  & 1.68      & 680       & 1.59    &  V10      \\
 110205 & 35 & 340  & 3.0      & 950  & 0.4 & 9.2   & 1.50      & 61        & 0.78    & C11,Z11,B12\\
 110213 & 24 & 90   & 1.7      & 340  & 1.1 & 4.1   & 0.75      & 2.4       & 0.83    & C11       \\
  \hline \hline 
\end{tabular}
\begin{minipage}{170mm}
\vspace*{2mm}
  (1) number of optical measurements,
  (2) epoch of first measurement,
  (3) power-law rise index of the best-fit {\sl for} $t_o \geq 0$ (uncertain because of its 
        degeneracy with the time-origin $t_o$),
  (4) epoch of light-curve peak (is usually close to the break-time $t_b$),
  (5) width of the peak at half maximum, in log scale, and for $t_o = 0$ ($fwhm$ is a substitute
        for the smoothness parameter $\xi$)
  (6) 2eV flux at light-curve peak (is usually within a factor 2 to the normalization factor $f$),
  (7) best-fit power-law decay index, 
  (8) epoch of last measurement used for fit,
  (9) reduced $\chi^2$ of the best-fit obtained with $t_o \geq 0$,
  (10) references for data.
\end{minipage}
\label{otable}
\end{table*}

 The decay index has an absolute error $\sigma(\alpha_d) < 0.05$ but the rise index error 
$\sigma(\alpha_r)$ is much larger because to its degeneracy with $t_o$, both quantities being 
constrained by the measurements during the light-curve rise. The rise index $\alpha_r$ decreases 
with increasing $t_o$, the uncertainties of $t_o$ and $\alpha_r$ being determined by how well the 
measurements during the flux rise can be described by a concave (holding water) "power-law" (for $t_o < 0$) 
or by a convex (not holding water) power-law (for $t_o > 0$). For most optical light-curves, there are 
only several measurements during the rise, thus a curvature in the power-law rise is often allowed, 
hence the uncertainties of $t_o$ and $\alpha_r$ are often substantial.

 For about 4/5 of the 31 optical peaks, there is no curvature in the optical flux power-law rise, 
in the sense that the best-fit obtained with $t_o = 0$ (afterglow begins at GRB trigger) is 
statistically as good as the best-fit obtained with a free $t_o$. More precisely, 
the relevance of $t_o$ as a fit parameter is established from the $F$-test probability that the 
increase in the best-fit $\chi^2$ resulting from fixing $t_o=0$ is accidental: we consider that 
$t_o \neq 0$ is required if the $F$-test probability is less than 10 percent, i.e. there is a 
higher than 90 percent chance that the improvement in the best-fit $\chi^2$ obtained with a free 
$t_o$ is real.

 The 1/5 of optical peaks for which the curvature of the afterglow rise requires $t_o \neq 0$ are listed
in Table \ref{totable}, together with the $\chi^2$ of the best fits obtained for a free $t_o$, for $t_o = 0$, 
and the $F$-test probability for the $t_o = 0$ fit to be accidentally worse than for the
$t_o \neq 0$ fit. There is only one afterglow (071010B) for which the optical flux rise is convex and 
requires $t_o < 0$ (afterglow begins before the GRB trigger); for the remaining five cases, $t_o > 0$ 
(afterglow begins after GRB trigger) is required by a concave light-curve rise. 

\begin{table*}
\caption{Optical afterglows for which $t_o \neq 0$ provides a better fit to the light-curve rise
     than $t_o = 0$, i.e. for which the afterglow beginning is not the GRB trigger}
\begin{tabular}{lcccccccccccccccc}
   \hline \hline
   GRB       &  $N_r$ & $\chi^2_\nu$ & $t_o^{(min)}$ & $t_o^{(max)}$ & $\chi^2_\nu$ & p (\%) \\
   afterglow &        & ($t_o\neq 0$)&   (s)         &  (s)          & ($t_o  = 0$) & ($t_o=0$)  \\
             &   (1)  &   (2)        &   (3)         &  (4)          &   (5)        & (6)   \\
   \hline
  060206     &  5     &  0.72        &  1690         &  1710         &  1.11        & 0.03\\
  061007     &  3     &  0.55        &   20          &   40          &  0.88        & 7.0 \\
  071010B    &  12    &  0.72        &  -400         &  -120         &  0.83        & 7.5 \\
  081008     &  7     &  2.30        &   80          &  110          &  3.51        & 0.90 \\
  100901(2)  &  9     &  1.59        &  11600        &  14200        &  2.17        & 2.4 \\
  110205     &  9     &  0.78        &   50          &   240         &  0.97        & 1.5 \\
  \hline \hline 
\end{tabular}
\begin{minipage}{85mm}
\vspace*{2mm}
 (1) number of measurements during the afterglow rise,
 (2) reduced $\chi^2$ of the best-fit obtained with free $t_o$,
 (3) lower limit of the 90 percent confidence level ($cl$) on $t_o$,
 (4) upper limit of the 90 percent $cl$ on $t_o$,
 (5) reduced $\chi^2$ of the best-fit obtained with $t_o =0$,
 (6) chance probability (in percent) that $t_o = 0$. 
\end{minipage}
\label{totable}
\end{table*}

 A second argument for the possibility that some optical afterglows begun after the GRB trigger is
based on the model interpretation of afterglow peaks. Numerical calculations of afterglow light-curves
in the two forward-shock models ("off jet-aperture observer" and "deceleration onset") that will be 
discussed below yield peaks with a $fwhm$ of at least 0.6 dex (factor 4 increase in time), 
depending mostly on the post-peak decay index. The effect of a shift $t_o > 0$ of the afterglow 
beginning relative to the GRB trigger is to increase the $fwhm$ of the peak if time were measured 
from $t_o$ instead of the GRB onset; thus, all peaks in Table \ref{otable} with $fwhm < 0.6$ would require 
$t_o > 0$ if they are to be accounted for the peak models considered below. There are five such 
narrow optical peaks in Table \ref{otable} and, for two of them (see Table \ref{totable}), we do find 
that $t_o > 0$ is also required by the best-fit to their rises. 

\begin{figure}
\centerline{\psfig{figure=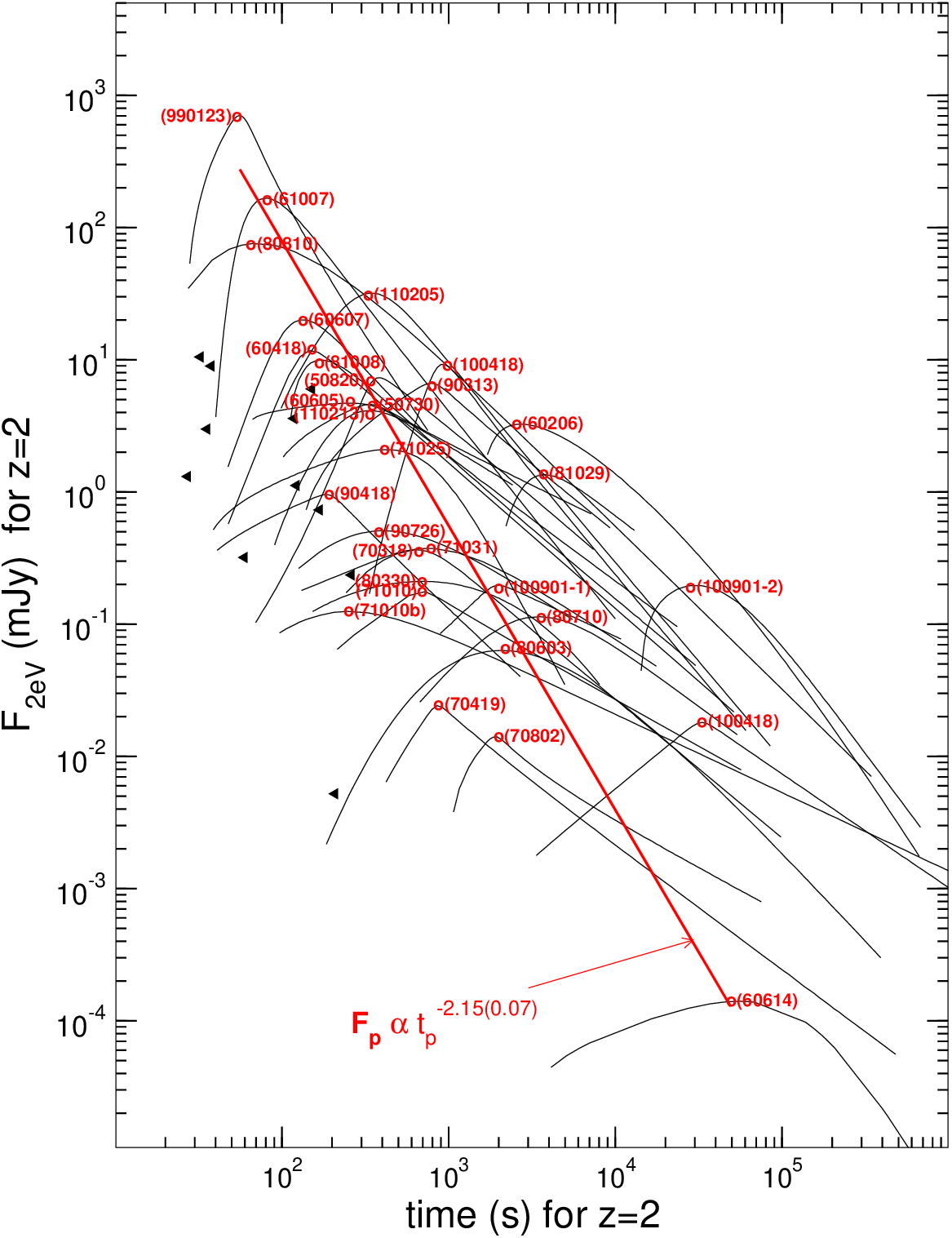,width=85mm}}
\caption{Smoothly broken power-law fits to 31 optical peaks.
   The peak location with time measured since the GRB trigger is labelled by the GRB name; 
   the best-fit in log-log scale is shown with a red line. 
   Triangles show the first measurement for several optical afterglows whose rises were missed,
   i.e. those peaks occurred before the first measurement. }
\label{opeaks}
\end{figure}

 Figure \ref{opeaks} shows the broken power-law fits to the 31 optical peaks and the best-fit power-law 
relation between peak time $t_p$ and peak flux $F_p$, obtained by calculating the coefficients $a$ and 
$b$ that minimize
\begin{equation}
 \chi^2 = \sum_i \frac { [ \log F_{p,i} - (a + b \log t_{p,i}) ]^2 } 
 {\sigma^2 (\log F_{p,i}) + b^2 \sigma^2 (\log t_{p,i})}
\end{equation}
For $t_o=0$ (i.e. the peaks labeled in Fig \ref{opeaks} by the GRB name), the best-fit is 
$F_p \propto t_p^{-2.15 \pm 0.07}$, assuming the same relative error $\epsilon (t_p) = \epsilon (F_p) 
= 0.15$ for all afterglows. The exponent changes by about 0.10 if either $\epsilon (t_p)$ or 
$\epsilon (F_p)$ is varied by a plausible 0.05, thus the full $1\sigma$ uncertainty of the $t_p$ 
exponent is about 0.12 (adding in quadrature the two uncertainties). 

 This best-fit is significantly less steep than the one previously reported by us (Panaitescu \& Vestrand 
2011), $F_p \propto t_p^{-3.2 \pm 0.2}$, for a set of 16 optical peaks, owing to the inclusion of 
a few afterglows which peak at later epochs. It is, however, consistent with the correlation found
by Liang et al (2012), $F_p \propto t_p^{-1.9 \pm 0.3}$, for a set of 39 afterglows. Twenty-three of the peaks 
in Table \ref{otable} are in Liang's sample, for the other 16 afterglows either the redshift is not known or 
the optical measurements do not allow a clear identification of a peak in the light-curve.

 The linear (Pearson) correlation coefficient 
\begin{equation}
 r (x,y) = \frac{<xy>-<x><y>}{\sigma_x \sigma_y}
\end{equation}
for the optical peaks shown in Figure \ref{opeaks} is  $r(\log F_p, \log t_p) = -0.77 \pm 0.01$. 
For 31 points, 
the corresponding probability that this correlation occurred by chance (i.e. in the null hypothesis, 
that the peak epoch is not correlated with the peak flux) is $\log p = - 6.5$. 

 If we allow for a time-shift $\tau = \frac{1}{2}[t_o^{(min)} + t_o^{(max)}]$ between the GRB trigger and 
the onset of the optical afterglow, with $[t_o^{(min)}-t_o^{(max)}]$ the 90 percent confidence level 
($cl$) on the afterglow onset epoch, determined by a variation of $\Delta \chi^2 = 2.7$ around the best-fit 
obtained with $t_o = 0$, then $r[\log F_p, \log (t_p-\tau)] = -0.80 \pm 0.02$ and the corresponding
chance probability $\log p = -6.9$ is 3 times lower than for the $F_p - t_p$ correlation. The new best-fit is 
\begin{equation}
  F_p^{(opt)} = 3.6 \times 10^5\, (t_p-\tau)^{-\gamma_o} \, {\rm (mJy)} \;, \gamma_o = 2.00 \pm 0.08
\label{optplaw}
\end{equation}
assuming a relative error $\epsilon(F_p) = 0.15$ for all afterglows and using the uncertainty 
of peak-time shift, $\sigma (\tau) \equiv 0.60 (t_o^{(max)}-t_o^{(min)})/2$, as the $1\sigma$ error for 
$t_p-\tau$. These results are almost unchanged if one uses $t_o$ of the best-fit instead of the middle 
of the 90 percent $cl$.

\section{X-ray peaks}

 X-ray afterglows light-curves displaying a peak are rarely observed by Swift-XRT, in the sense that we
found only such 14 cases among the several hundred of X-ray light-curves in the UK Science Data Center
repository. One reason for that rarity is that the prompt emission and its tail overshine the afterglow
emission until tens--hundreds of seconds after trigger. Another reason is that we have retained only
afterglows whose emergence after the GRB tail was sufficiently well sampled. A strong spectral hardening
is usually observed at the emergence of the X-ray afterglow, but such a spectral evolution was not a 
criterion for selecting the X-ray afterglows. 

 Some of the X-ray light-curve peaks are displayed by the 0.3-10 keV band light-curves found at the Swift-XRT
light-curve repository (Evans et al 2007, 2009, 2010), but others display a peak only in the XRT processed
10 keV light-curve (for all, we have calculated the 1 keV flux using the X-ray spectral slope reported at
the Swift-XRT spectrum repository -- Evans et al 2007). The reason for which some 0.3-10 keV light-curves
do not display a peak although the processed 10 keV light-curves do so is that, just after the emergence 
of the afterglow rise, the {\sl softer} GRB tail emission contributes more to the 0.3-10 keV flux than to 
the 10 keV emission, relative to the {\sl harder} afterglow flux. Thus, for afterglows peaking shortly after 
their emergence from under the GRB tail, the rise of the 0.3-10 keV afterglow flux can be masked by the GRB 
tail. Whenever they displayed a clear peak, we used the 0.3-10 keV light-curves instead of the 10 keV fluxes, 
because the flux errors of the 10 keV processed light-curves are underestimated, as the errors of the 
interpolated hardness ratios (or spectral slopes) were not taken into account in the calculation of the 
10 keV flux from the 0.3-10 keV count rate.

\begin{figure*}
\centerline{\psfig{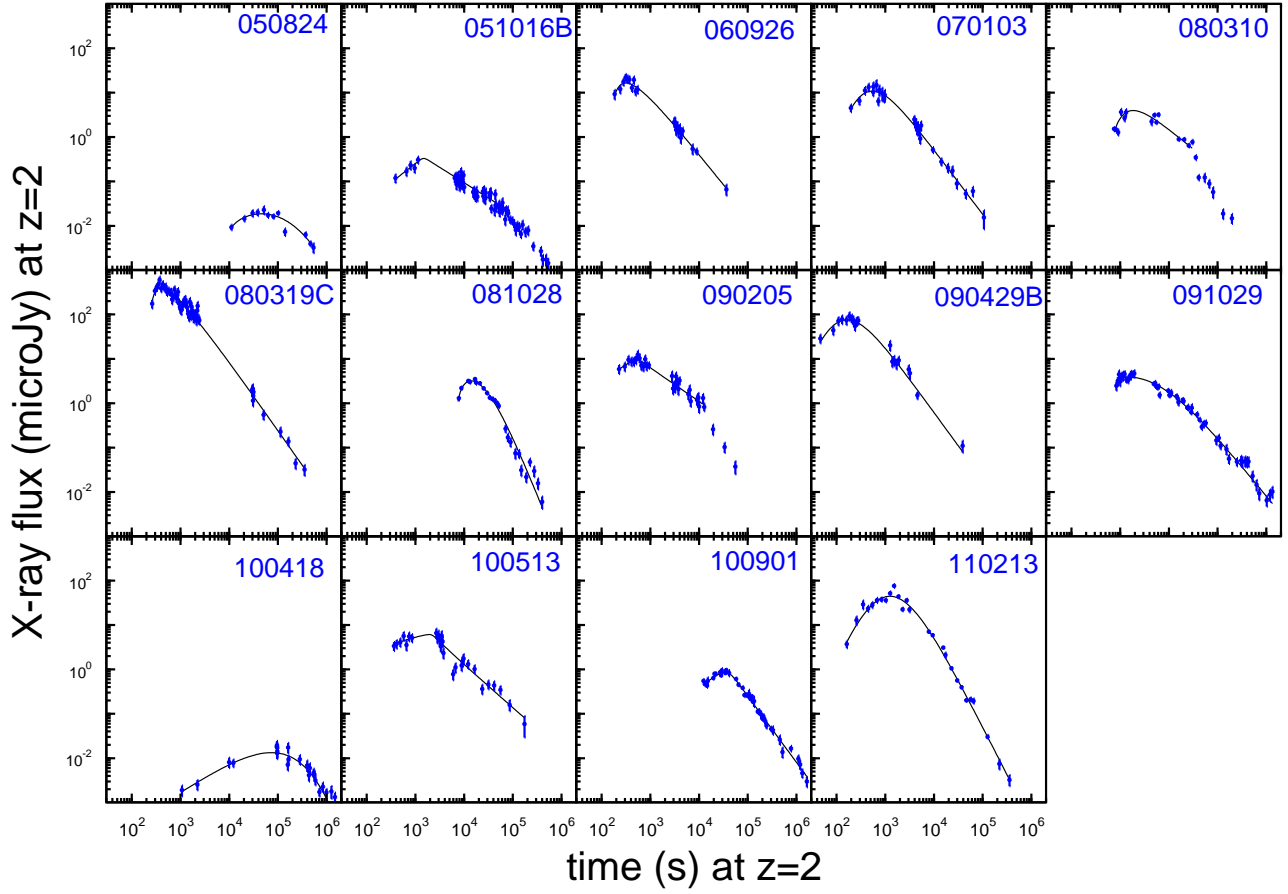}}
\caption{ Light-curves of 14 X-ray afterglows with peaks and broken power-law fits to those peaks. }
\label{allx}
\end{figure*}

 The 14 X-ray afterglows with peaks are shown in Figure \ref{allx} and some of the parameters obtained with 
the smoothly broken power-law fit are listed in Table \ref{xtable}. 
The allowed range for the X-ray emission onset epoch $t_o$ were calculated only 
for $t_o > 0$, the true 90 percent $cl$ being even larger, owing to the ($\alpha_r,t_o$) fitting degeneracy.
For most of X-ray afterglows, the 90 percent $cl$ on $t_o$ extends from trigger to near the first measurement, 
with only two exceptions: for 080319C, $t_o > 220$ s, while for 091029, $t_o > 410$ s, the probability 
(calculated from the corresponding $\chi^2$) that $t_o = 0$ being around 10 percent for both afterglows.

\begin{table*}
\caption{Parameters of the smoothed broken power-law best-fit to 14 {\sl X-ray} light-curves with peaks 
   (Fig \ref{allx}), at $z=2$, and other quantities of interest describing those light-curves, as defined 
   in Table \ref{otable}. Last column gives the slope of the X-ray spectrum $F_\nu \propto \nu^{-\beta_x}$ 
    (and its 90 percent $cl$). }
\begin{tabular}{lcccccccccccccccc}
   \hline \hline
 GRB    & N & $t_1$ &$\alpha_r$& $t_p$ &$fwhm$& $F_p$ &$\alpha_d$& $t_{end}$ &$\chi^2_\nu$& $\beta_x$ \\
afterglow & &  (s)  &          &  (s)  & (dex)&($\mu Jy$)&        &    (ks)   &                  &        \\
  \hline
 050824 & 10& 11000 &  0.9    & 47000 & 1.1  &  0.019 &  4.47    &   540     &  0.64  & 0.95(.18) \\
 051016B& 46&  390  &  0.8    &  1500 & 0.9  &  0.32  &  0.74    &   60      &  0.85  & 0.89(.13) \\
 060926 & 21&  185  &  4.0    &  310  & 0.6  &  17    &  1.36    &   36      &  0.90  & 1.00(.29) \\
 070103 & 27&  200  &  3.6    &  540  & 1.0  &  11    &  1.45    &   106     &  0.87  & 1.31(.24) \\
 080310 & 12&  740  &  1.1    & 2000  & 0.9  &  4.5   &  0.75    &   30      &  3.41  & 1.09(.07) \\
 080319C& 58&  260  &  1.8    &  350  & 0.6  &  430   &  1.82    &   350     &  1.08  & 0.94(.32) \\
 081028 & 19& 7800  &  0.4    & 14600 & 0.6  &  3.3   &  2.31    &   400     &  0.94  & 1.03(.07) \\
 090205 & 31&  230  &  1.1    &  490  & 0.8  &  9.3   &  0.78    &   13      &  0.80  & 1.03(.15) \\
 090429B& 20&   46  &  1.3    &  180  & 1.0  &  75    &  1.44    &   39      &  1.08  & 1.00(.24) \\
 091029 & 53&  830  &  0.2    & 1500  & 1.0  &  3.4   &  1.34    &  1300     &  0.98  & 1.09(.10) \\
 100418 & 26& 1070  &  0.7    & 76000 & 1.6  &  0.013 &  3.5     &  3700     &  1.59  & 1.27(.40) \\
 100513 & 22&  360  &  0.2    & 1900  & 1.1  &  6.1   &  0.96    &   170     &  1.11  & 1.27(.27) \\
 100901 & 42& 12000 &  0.7    & 34300 & 0.8  &  0.87  &  1.49    &  1640     &  1.63  & 1.09(.07) \\
 110213 & 20&  160  &  0.9    & 1300  & 1.0  &  44    &  2.09    &   360     &  1.66  & 0.99(.07) \\
  \hline \hline 
\end{tabular}
\label{xtable}
\end{table*}

 The 90 percent $cl$s on $t_o$ are often quite large, being in some cases the entire interval $[0,t_1]$
from GRB trigger and until first measurement, because of the degeneracy between $t_o$ and the rise slope
$\alpha_r$. Narrower ranges for $t_o$ are obtained if $\alpha_r$ is restricted to, for instance, the
range $[2,3]$ spanned by the expectations for the pre-deceleration forward-shock model (and for homogeneous
ejecta). However, because some peaks display a rise slower than $t^2$ if time is measured 
from GRB trigger, it is necessary to allow $t_o < 0$, so that the rise of such afterglows becomes faster 
when time is measured from $t_o$. With the time shift $\tau$ determined as above, but using the 90 percent
$cl$s on $t_o$ calculated from fits with the $2 < \alpha_r < 3$ restriction, the $F_p - t_p$ correlation
is significantly stronger ($\log p = -7.7$) and the best-fit is $F_p \propto t_p^{-1.90 \pm 0.07}$.

 The last column of Table \ref{xtable} lists the slope of the afterglow X-ray spectrum reported at the Swift-XRT 
spectrum repository. The error-weighted average of those 14 slopes is
\begin{equation}
 (F_\nu \propto \nu^{-\beta_x}) \;\; \overline{\beta_x} = 1.04 \pm 0.02 
\label{bx}
\end{equation}
error being $1\sigma$, and with a $\chi^2_\nu = 1.13$ for all slopes being consistent with the average 
$\overline{\beta_x}$ given above. 

\begin{figure}
\centerline{\psfig{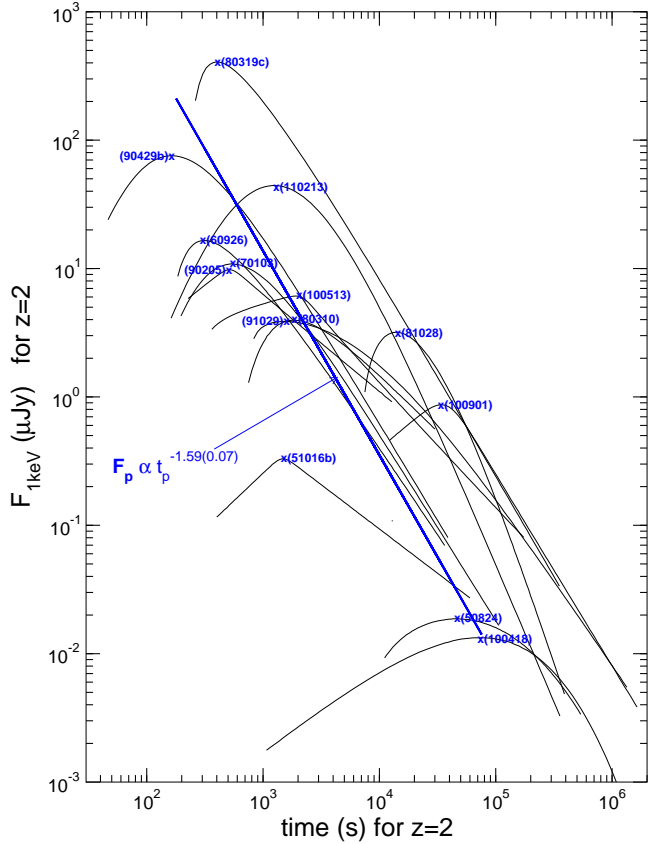}}
\caption{Broken power-law fits to the peaks of 14 X-ray afterglows and the power-law fit to their
       peak fluxes and epochs (blue line). Peak location is indicated by the afterglow name. }
\label{xpeaks}
\end{figure}

 Figure \ref{xpeaks} shows the broken power-law fits to the 14 X-ray peaks together with the best-fit 
to their peak fluxes and epochs. For $t_o =o$, the linear correlation coefficient of the peak fluxes and peak
epochs is $r(\log F_p, \log t_p) = -0.83 \pm 0.01$, chance probability $\log p = -3.7$, and the best-fit
to the peak locations is $F_p \propto t_p^{-1.59 \pm 0.07}$. For the peak epochs shifted by the center 
of the 90 percent $cl$ on $t_o >0$, we obtain $r[\log F_p, \log (t_p -\tau)] = -0.85 \pm 0.02$, 
$\log p = -3.9$ and
\begin{equation}
 F_p^{(xray)} = 740\; (t_p -\tau)^{-\gamma_x}\; {\rm (mJy)} \;,\; \gamma_x = 1.63 \pm 0.09
\label{xplaw}
\end{equation}
which is shallower than for optical light-curves (eq \ref{optplaw}):
\begin{equation}
 \gamma_x - \gamma_o = -0.37 \pm 0.12 \;.
\label{gox}
\end{equation}

\begin{figure}
\centerline{\psfig{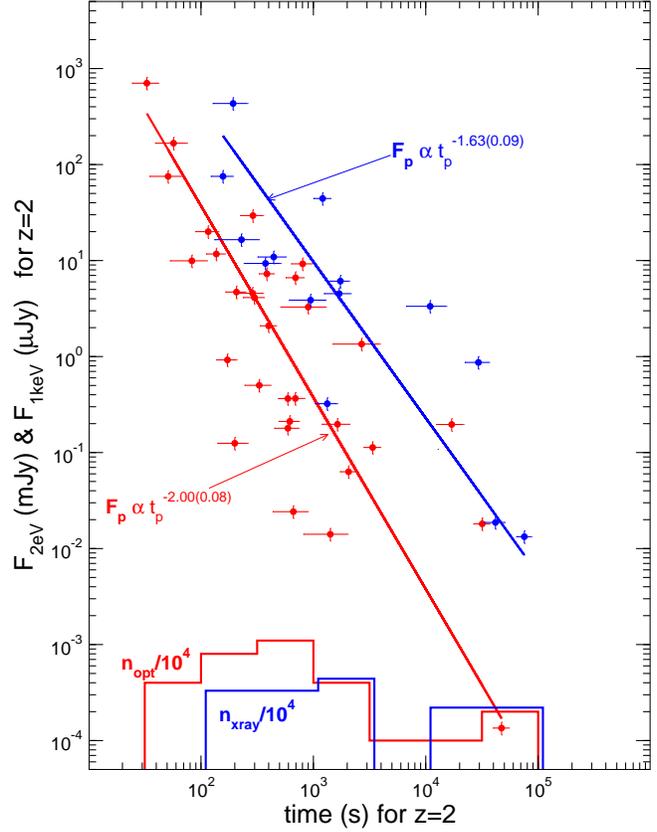}}
\caption{ 31 optical peaks (red symbols) and 14 X-ray peaks (blue) with the peak epoch measured
   from the middle value of the 90 percent confidence level on $t_o$, determined by fitting light-curves
   with the function in equation (\ref{fit}). 
   The histograms at the bottom show the distributions of optical (red) and X-ray (blue) peak epochs. }
\label{oxpeaks}
\end{figure}

 To compare the distributions of optical and X-ray peak epochs, one can calculate 
\begin{equation}
 \chi^2 = \sum_i \frac{( \sqrt{N_x/N_o} n_{i,o} - \sqrt{N_o/N_x} n_{i,x} )^2}{n_{i,o} + n_{i,x}}
\end{equation}
with $n_i$ the number of optical or X-ray peaks in time-bin $i$ and $N$ the total number of optical or X-ray
afterglows. For our two samples, the resulting $\chi^2 = 6.7$ for 7 $dof$ corresponds to a 50 percent probability
that the optical and X-ray peaks shown in Figure \ref{oxpeaks} are drawn from the same distribution. 

\vspace*{4mm}
\centerline{\bf ORIGIN OF PEAK FLUX -- }
\centerline{\bf PEAK EPOCH ANTICORRELATION}
\vspace*{2mm}

 As indicated in Figure \ref{opeaks}, for some optical afterglows, we have seen a decaying flux since first 
measurement,
thus the peak of the optical light-curve occurred at an earlier time than for the afterglows listed in Table 
\ref{otable}.
This means that equation (\ref{optplaw}) represents only the bright/late edge of the distribution of all
optical light-curve peaks in the $(F_p,t_p)$ plane. 

 There are two mechanisms that can produce a peak in
the afterglow light-curve: $1)$ an observer location that is (for some time) outside the aperture of the 
relativistic afterglow jet, such that the received afterglow flux rises when the observer is still outside
the $1/\Gamma$ aperture of the afterglow emission, $\Gamma$ being the ever-decreasing jet Lorentz factor,
and $2)$ a dynamics of the afterglow-producing shock such that the flux of that shock's emission rises for
some time. 

 In the former case, the peak of the afterglow light-curve occurs when, owing to the gradual increase of 
the $\Gamma^{-1}$ aperture of the Doppler-boosted emission cone, the center of the jet becomes visible to 
the observer. In the latter case, the afterglow rises as more energy is added to the afterglow shock, 
the light-curve peak corresponding to a change in the rate at which that energy is injected into the shock.
A sudden change in the injected power occurs naturally if the ejecta are contained in a well-defined shell, 
so that the dynamics of both shocked fluids (ejecta and swept-up medium) changes (i.e. deceleration sets in) 
when the reverse shock crosses the entire ejecta shell.

 For either model for light-curve peaks (off-jet observer location or onset of deceleration), we will search
for the parameter whose variation among afterglows can explain the existence of the bright/late edge of
peak fluxes and epochs in the $(F_p,t_p)$ plane (i.e. can induce the $F_p - t_p$ correlations of eqs
\ref{optplaw} and \ref{xplaw}), and that can yield a steeper $F_p - t_p$ dependence for optical peaks than
for X-ray peaks.

\section{Off-jet observer location}

 A structured or dual outflow model was first proposed by Berger et al (2003) to explain the radio emission 
of GRB afterglow 030329. The existence of many afterglows with decoupled optical and X-ray afterglow 
light-curves (i.e. with different decay rates at these two frequencies or with chromatic X-ray breaks)
has revived interest in this model (e.g. Racusin et al 2008). 

 The basic feature of this model for light-curve peaks is that the observer is located outside the opening 
of the afterglow jet, $\theta_j$, at an angle $\theta_o > \theta_j$. That is a somewhat unappealing feature, 
as it implies that the GRB emission is produced by a different jet, whose aperture includes the direction 
toward the observer, allowing us to localize the burst and follow its afterglow. A more palatable version 
of it is an outflow with an non-uniform angular structure (i.e. distribution of ejecta energy with direction 
in the jet), such as a bright core moving toward the observer and emitting $\gamma$-rays during the prompt 
phase and an envelope surrounding the core an dominating the afterglow emission at later times. 

 For simplicity of derivations, we focus below on the off-aperture jet, but the results for the peak flux 
and epoch should be the same for an envelope outflow.
 The emission received by the observer is that in the comoving frame, boosted relativistically, 
$F_\nu = {\cal D}^3 F'_{\nu/{\cal D}}$, where ${\cal D}$ is the relativistic boost factor 
\begin{equation}
 {\cal D} = \frac{1}{\Gamma (1- v \cos \theta_o)} \simeq \frac{2 \Gamma}{\Gamma^2 \theta_o^2 +1}
          \simeq \left\{ \begin{array}{ll} 
        \hspace*{-2mm} 2/(\Gamma\theta_o^2) & \hspace*{-2mm} (\Gamma > \theta_o^{-1}) \\
        \hspace*{-2mm} 2 \Gamma             & \hspace*{-2mm} (\Gamma < \theta_o^{-1}) \end{array} \right.
\end{equation}
$\Gamma$ being the jet Lorentz factor, and assuming $\Gamma \gg 1$ and $\theta_o \ll 1$. 
At early times (i.e. when $\Gamma > \theta_o^{-1}$), the Doppler 
boost increases while at later time (i.e. when $\Gamma < \theta_o^{-1}$), the Doppler factor decreases,
both owing to the jet deceleration. The behaviour of that factor is convolved with the decrease of the
afterglow flux in the comoving frame, at frequencies above the peak of the comoving-frame (synchrotron) 
emission spectrum, also due to the continuous jet deceleration. For $\Gamma > \theta_o^{-1}$, the
increase of ${\cal D}$ dominates and the observer sees an increasing afterglow flux, until $\Gamma =
 \theta_o^{-1}$, when the afterglow light-curve peaks, being followed by a decreasing flux when 
$\Gamma < \theta_o^{-1}$. Thus the afterglow peak occurs when the ever-widening $\Gamma^{-1}$-opening 
cone of relativistically-boosted emission contains the direction toward the observer, at which time the 
Doppler boost is 
\begin{equation}
 {\cal D} (t_p) = \Gamma (t_p) = \theta_o^{-1} \;.
\label{D}
\end{equation}

 The observer offset angle $\theta_o$ is the only parameter to which makes sense to attribute the
the $F_p-t_p$ correlation, as an increasing angle $\theta_o$ yields a later peak epoch $t_p$ and a
lower peak flux $F_p$ (see below). To calculate the $F_p-t_p$ relation induced by a variable angle
$\theta_o$, one has to relate $F_p$ and $t_p$ to the offset angle. 
The observer-frame arrival time $t$ of the photons emitted by a relativistic source moving at Lorentz 
factor $\Gamma(r)$ and angle $\theta_o$ relative to the direction toward the observer is found by 
integrating 
\begin{equation}
  cdt = \frac{dr}{v} - dr \cos \theta_o \simeq \frac{dr}{2} (\Gamma^{-2} + \theta_o^2) \;.
\label{dt}
\end{equation}

 To calculate the afterglow flux at $t_p$, we restrict our attention to the forward-shock synchrotron
emission (i.e. the emission from the shocked ambient medium), and leave the derivation of the $F_p-t_p$
relation for the reverse-shock emission to the fans of that model. In the comoving frame, the
forward-shock synchrotron emission peaks at frequency 
\begin{equation}
 \nu'_i \propto \gamma_i^2 B \propto \Gamma^3 n^{1/2}
\label{nui}
\end{equation}
with $\gamma_i \propto \Gamma$ being the typical post-shock electron energy, $B^2 \propto \Gamma n'
\propto \Gamma^2 n$ the magnetic field energy density, $n' \propto \Gamma n$ the post-shock particle
density, and $n$ the density of the circumburst medium. 
The comoving frame synchrotron flux density at $\nu'_i$ is 
\begin{equation}
 F'_{\nu'_i} \equiv F'_o \propto N_e B \propto M \Gamma n^{1/2}
\end{equation}
where $N_e \propto M$ is the number of forward-shock electrons and $M$ the mass of the energized ambient
medium. The synchrotron spectrum has a "cooling" break at the characteristic synchrotron frequency 
$\nu'_c$ of the electrons whose radiative cooling timescale ($t'_{rad} = E'_{el}/P'_{syn} \propto 
\gamma_c/ (\gamma_c^2 B^2) = \gamma_c^{-1} B^{-2}$) equals the dynamical timescale ($t'_{dyn} \propto r/\Gamma$),
thus $\gamma_c \propto \Gamma B^{-2} r^{-1}$ and 
\begin{equation}
 \nu'_c \propto \gamma_c^2 B \propto \Gamma^2 B^{-3} r^{-2} \propto \Gamma^{-1} n^{-3/2} r^{-2} \;.
\label{nuc}
\end{equation}

 In the observer frame, the characteristics of the synchrotron spectrum are
\begin{equation}
 F_o = {\cal D}^3 F'_o \quad , \quad \nu_i = {\cal D} \nu'_i \quad , \quad \nu_c = {\cal D} \nu'_c 
\label{Fo}
\end{equation}
and the afterglow flux at an observing frequency $\nu$ is
\begin{equation}
  F_\nu (\nu_c < \nu < \nu_i) = F_o \left( \frac{\nu_c}{\nu} \right)^{1/2} 
                                \propto {\cal D}^{7/2} F'_o (\nu'_c)^{1/2} 
\label{F1}
\end{equation}
\begin{equation}
  F_\nu (\nu_i < \nu < \nu_c) = F_o \left( \frac{\nu_i}{\nu} \right)^{\beta} 
                                \propto {\cal D}^{3+\beta} F'_o (\nu'_i)^\beta 
\end{equation}
\begin{equation}
  F_\nu (\nu_i,\nu_c < \nu)  = F_o \left( \frac{\nu_i}{\nu_c} \right)^{\hspace*{-1mm}\beta-1/2} 
                      \hspace*{-1mm}  \left( \frac{\nu_c}{\nu} \right)^{\beta} 
                       \hspace*{-1mm} \propto \hspace*{-1mm}
        {\cal D}^{3+\beta} F'_o (\nu'_i)^\beta  \hspace*{-1mm} \left( \frac{\nu'_c}{\nu'_i} \right)^{1/2} 
\label{F3}
\end{equation}
where $\beta$ is the slope of the afterglow spectrum at frequency observing $\nu$.
The case given in equation (\ref{F3}) applies only to optical light-curves because the implied
spectrum ($F_\nu \propto \nu^{-1/2}$) could be compatible with that of optical afterglows but
is clearly harder than observed for X-ray afterglows (eq \ref{bx}).

\subsection{Synchrotron forward-shock emission from conical jets}

 For an adiabatic jet that does not spread laterally (because, for instance, the jet is embedded in 
a confining envelope), conservation of energy during the interaction between the jet and the ambient 
medium reads $E \propto \Gamma^2 M$ with $E$ the jet energy and $M \propto n(r) r^3 \propto r^{3-s}$.
Then, for a $n \propto r^{-s}$ (with $s < 3$) radial distribution of the ambient medium density, 
the jet dynamics is given by
\begin{equation}
  \Gamma \propto r^{-(3-s)/2}
\label{Gamma}
\end{equation}
and equation (\ref{dt}) yields
\begin{equation}
 t = \frac{r}{2} \left( \frac{1}{4-s} \Gamma^{-2} + \theta_o^2\right) 
\end{equation}
hence the epoch of the light-curve peak, $t_p \equiv t|_{\Gamma = \theta_o^{-1}}$, satisfies 
$t_p \propto r_p \theta_o^2$, where $r_p \equiv r|_{\Gamma = \theta_o^{-1}} \propto \theta_o^{2/(3-s)}$
is the jet radius at which the light-curve peaks, the last relation following from equation (\ref{Gamma}). 
Thus 
\begin{equation}
 t_p \propto \theta_o^{(8-2s)/(3-s)} \;.
\label{tptheta}
\end{equation}
Eliminating $\theta_o$ between the expressions for $r_p$ and $t_p$, it follows that 
\begin{equation}
 r_p \propto t_p^{1/(4-s)} \;.
\label{rp}
\end{equation}
 
 Substituting $n(r_p)$, $M(r_p)$, and $\Gamma(r_p)$ in equations (\ref{nui}) -- (\ref{nuc}) and using
equation (\ref{rp}), we get
\begin{equation}
 F'_o (t_p) \propto r_p^{3/2-s} \;, \quad \nu'_i (t_p) \propto r_p^{s-9/2} \;, 
 \quad \nu'_c (t_p) \propto r_p^{s-1/2}  \;.
\label{tp0}
\end{equation}
The dependence of the afterglow peak flux $F_p \equiv F_\nu (t_p)$ on the light-curve peak epoch $t_p$
can be now calculated from equations (\ref{Fo})--(\ref{F3}), with ${\cal D} \propto \theta_o^{-1} \propto
t_p^{(s-3)/(8-2s)}$, and using the spectral characteristics given in equation (\ref{tp0}), with 
$r_p(t_p)$ from equation (\ref{rp}). 

\subsubsection{Homogeneous medium (s=0)} 

 With the above-mentioned substitutions, one obtains the following exponents for the $F_p \propto 
t_p^{-\gamma}$ correlation for afterglow light-curves from off-aperture jets interacting with a 
homogeneous medium ($s=0$): 
$\gamma_\theta (\nu_c < \nu < \nu_i) = 1$, which is too small compared with the values measured for 
optical and X-ray peaks, and 
\begin{equation}
  \nu_i < \nu < \nu_c:\; \gamma_\theta = \frac{3}{4} (2\beta + 1) \; , \;
  \nu_i,\nu_c < \nu:\; \gamma_\theta = \frac{1}{4} (6\beta + 1) \;.
\label{oxs0}
\end{equation}
For the former case, the average measured X-ray spectral slope (eq \ref{bx}) implies that $\gamma_x = 2.3$, 
which is too large than what is measured for X-ray peaks (eq \ref{xplaw}). The second case leads to 
$\gamma_x = 1.81 \pm 0.03$, which is almost compatible with observations. However, this case cannot 
account for the slope $\gamma_o$ measured for X-ray peaks, {\sl if} the optical and X-ray 
afterglow emissions arose from the same outflow (same electron population): for $\nu_i, \nu_c < 
\nu_o$ (i.e. $\beta_o = \beta_x$), we expect $\gamma_o = \gamma_x$, which is inconsistent with
observations; for $\nu_i < \nu_o < \nu_c < \nu_x$, (i.e. $\beta_x = \beta_o + 1/2$), equation 
(\ref{oxs0}) implies that $\gamma_x - \gamma_o = 1.5(\beta_x - \beta_o) - 0.5 = 1/4$, contrary 
to observations (eq \ref{gox}). 

\subsubsection{Wind-like medium (s=2)} 

 Similar to above, for a medium having the $n \propto r^{-2}$ radial stratification expected for the
winds of massive stars (as the progenitors of long GRBs), we obtain $\gamma_\theta (\nu_c < \nu < \nu_i) = 0.75$, which is smaller than measured for optical and X-ray peaks, and
\begin{equation}
  \nu_i < \nu < \nu_c:\; \gamma_\theta = \frac{3}{2} \beta + 1 \; , \quad
  \nu_i,\nu_c < \nu:\; \gamma_\theta = \frac{3}{2} \beta  \;.
\label{oxs2}
\end{equation}
In the former case, the average X-ray spectral slope leads to $\gamma_x = 2.6$, which is larger
than observed, while the latter case leads to $\gamma_x = 1.56 \pm 0.03$, which is consistent with 
observations of X-ray peaks, but too small for the optical peaks slope. However, if $\nu_i < \nu_o < 
\nu_c < \nu_x$, then equation (\ref{oxs2}) yields $\gamma_x - \gamma_o = 1.5 (\beta_x - \beta_o) - 1 = 
-1/4$, which is marginally consistent with the measured value (eq \ref{gox}). Thus, a jet seen 
off aperture, interacting with a wind-like medium, can account for both the optical and X-ray peak
slopes provided that the synchrotron cooling frequency is between optical and X-rays.

\subsubsection{Jet half-opening and observer offset angle}

 In deriving the above scalings for the flux, we have ignored any intrinsic afterglow parameter and kept 
only the dependencies on the observer's offset angle $\theta_o$ because we are interested in the $F_p-t_p$ 
relation induced by the variation of $\theta_o$. The ignored afterglow parameters, kinetic energy per 
solid angle $E$ and ambient density $n$, are not the same for all afterglows, and their variation of $E$ 
and $n$ among afterglows yields a scatter in the $(F_p,t_p)$ the plane, along the relation induced by 
the variation of $\theta_o$.

 Equation (\ref{tptheta}) implies that, aside from that scatter induced by the variation of $E$ and $n$, 
the offset angle $\theta_o$ must span a range of 1 dex, to account for the 3 dex spread in observed peak 
time $t_p$ (Fig \ref{oxpeaks}). For typical afterglow energies and ambient medium densities, the jet
Lorentz factor at the earlier peak epochs shown in Figure \ref{oxpeaks}, of about 100 s, is expected
to be $\Gamma \sim 100$, which implies that the smallest offset is $\theta_o^{(min)} \sim 0.5\deg$.
To obtain a light-curve peak, the jet opening must be smaller than the observer's offset, thus the 
earliest peaks correspond to the narrowest jets, with $\theta_j^{(min)} < 0.5\deg$. If all jets
had the same opening, then the largest offset angle would be $\theta_o^{(max)} \sim 5 \deg$. 
However, the latest occurring peaks are not necessarily from narrow jets seen at a larger offset angle,
they could also be from wider jets: at 10 ks, the jet Lorentz factor is $\Gamma \sim 10$, hence the 
offset angle is $\theta_o \sim 5\deg$ and the widest jet could have $\theta_j^{(max)} < 5\deg$.
Therefore, the range of peak epochs shown in Figure \ref{oxpeaks} suggests that afterglow jets have
half-openings $\theta_j$ from 1/2 to 5 degrees and are seen at an angle $\theta_o = (1 \div N) \theta_j$,
with $N < 10$.

\subsection{Synchrotron forward-shock emission from a spreading jet interacting with a homogeneous medium}

 If the afterglow jet spreads laterally unimpeded, then the jet dynamics changes when the lateral spreading
(at the sound speed) increases significantly the jet opening, leading to other dependencies of the
peak flux $F_p$ on the peak epoch $t_p$ than derived above for a conical jet. 
Defining $r_j$ to be the radius at which the jet Lorentz factor has decreased to $\Gamma_j \equiv 
\theta_j^{-1}$, where $\theta_j$ is the initial jet half-angle, the jet dynamics is given by 
(Panaitescu \& Vestrand 2011)
\begin{equation}
  \Gamma (r) = \left\{ \begin{array}{ll}  \Gamma_j (r/r_j)^{-3/2}  & (r < r_j) \\
       \Gamma_j  e^{-(\frac{r}{r_j}-1)} (r/r_j)^{-1} & (r > r_j) \\ \end{array}  \right.
\label{jetdyn}
\end{equation}
for a homogeneous medium (the jet dynamical equations cannot be solved analytically for a wind-like medium). 
At $r < r_j$, the lateral spreading is negligible and the jet dynamics is the same as for a conical jet but 
at $r > r_j$ the lateral expansion of the jet more than doubled its initial aperture and an exponential 
deceleration sets in (Rhoads 1999). 

 With the exception of observer locations just outside the jet opening, the afterglow light-curve peak 
occurs after $r_j$ (i.e. $r_p > r_j$), when the jet lateral spreading is significant.
Equation (\ref{dt}) can be integrated first over the power-law jet deceleration and then over the exponential 
deceleration, yielding
\begin{equation}
 t = \frac{1}{2} r \theta_o^2 + \frac{r_j}{4\Gamma^2} \left[ 1 - \frac{r_j}{r} + \frac{1}{2}
             \left( \frac{r_j}{r}\right)^2 \right] \;.
\end{equation}
For $\theta_o \gg \theta_j$, the jet becomes visible to the observer at radius $r_p \gg r_j$, thus
\begin{equation}
 t_p \simeq \frac{1}{2} r_p \theta_o^2 + \frac{r_j}{4\Gamma_p^2} = 
             \left( \frac{r_p}{2} + \frac{r_j}{4} \right) \theta_o^2 \simeq \frac{1}{2} r_p \theta_o^2
\label{tp}
\end{equation}
where we used the definition of the afterglow peak time: $\Gamma (t_p) = \theta_o^{-1}$. 
That condition and the dynamics of the jet at $r > r_j$ (eq \ref{jetdyn}), 
imply that
\begin{equation}
 \frac{r_p}{r_j} e^{\frac{r_p}{r_j} -1 } = \Gamma_j \theta_o = \frac{\theta_o}{\theta_j} \;.
\end{equation}
This result can be simplified if it is assumed that all jets have the same $\theta_j$ and $r_j$,
which is equivalent to assuming that all jets have the same $\theta_j$ and ratio $E/n$. 
This simplification is motivated by that we are searching for the $F_p - t_p$ dependence arising only 
from varying the observer location, and not from varying the jet dynamical parameters (which is the 
subject of the next section, but in an other model for light-curve peaks). 
Thus,
\begin{equation}
 r_p e^{r_p/r_j} \propto \theta_o \propto (t_p/r_p)^{1/2}
\label{RP}
\end{equation}
the last scaling resulting from equation (\ref{tp}).

 It can be shown that, for a homogeneous medium, $M \propto r^2 \exp(2r/r_j)$ at $r > r_j$.
Substituting it and the jet dynamics (eq \ref{jetdyn}) in equations (\ref{nui}) -- (\ref{nuc}), 
we get
\begin{equation}
 F'_o (t_p) \propto r_p e^{r_p/r_j} \;,\; 
\end{equation}
\begin{equation}
 \nu'_i (t_p) \propto r_p^{-3} e^{-3r_p/r_j}  \;,\; 
\end{equation}
\begin{equation}
 \nu'_c (t_p) \propto r_p^{-1} e^{r_p/r_j} \;.
\end{equation}
From here, one can calculate $F_p \equiv F_\nu(t_p)$ by using equations (\ref{F1}) -- (\ref{F3}), 
with ${\cal D} \propto \theta_o^{-1}$ and by substituting $\exp(r_p/r_j)$ with the aid of equation 
(\ref{RP}):
\begin{equation}
  F_p (\nu_c < \nu < \nu_i) \propto t_p^{-1}
\end{equation}
\begin{equation}
  F_p (\nu_i < \nu < \nu_c) \propto r_p^{2\beta+1} t_p^{-(2\beta+1)}
\end{equation}
\begin{equation}
  F_p (\nu_i,\nu_c < \nu) \propto r_p^{2\beta-1} t_p^{-2\beta}
\end{equation}
Ignoring the $r_p$ factors (because $r_p$ has a weak, sub-logarithmic dependence on $t_p$, according
to eq \ref{RP}), one obtains:
$\gamma_\theta (\nu_c < \nu < \nu_i) = 1$, which is inconsistent with the $F_p-t_p$ correlation measured
for optical and X-ray peaks, and
\begin{equation}
  \nu_i < \nu < \nu_c:\, \gamma_\theta = 2\beta + 1 \; , \quad
  \nu_i,\nu_c < \nu: \; \gamma_\theta = 2 \beta \;.
\label{oxjet}
\end{equation}
 In the former case, the average X-ray spectral slope leads to an exponent $\gamma = 3.1$, inconsistent
with that measured for X-ray peaks, while the latter case yields $\gamma = 2.08 \pm 0.04$, which is too 
large for X-ray peaks but consistent with that of optical peaks, thus this case can explain the the
$F_p - t_p$ relation for optical peaks if optical were above the cooling frequency. If the cooling 
break were always between optical and X-rays, then the spreading-jet model expectation is $\gamma_x - 
\gamma_o = 2(\beta_x - \beta_o) - 1 = 0$, inconsistent with observations.

\section{Onset of deceleration}
\label{predec}

 In this model, for afterglow peak occurs when the reverse shock has finished crossing the ejecta
shell and the motion of the shocked fluids starts to decelerate (or to decelerate faster than before).
We focus only on the forward-shock emission, which depends on fewer afterglow parameters than
the reverse-shock's. Depending on the initial geometrical thickness $\Delta$ of the ejecta shell, 
the observer-frame deceleration timescale (i.e. the afterglow peak epoch) is either $\Delta/c$ 
(thick ejecta shell/relativistic reverse-shock case) or is determined by the parameter set 
$(E,\Gamma_0,n)$ (thin ejecta shell/semi-relativistic reverse-shock), where $E$ is the ejecta 
initial kinetic energy per solid angle and $\Gamma_0$ the ejecta initial Lorentz factor.

\subsection{Forward-shock synchrotron emission for a semi-relativistic reverse-shock}

 If the ejecta is shell is sufficiently thin (or, equivalently, sufficiently dense), the reverse-shock 
is semi-relativistic, the shocked fluids (ejecta and ambient medium) move at a Lorentz factor $\Gamma$
lower than but close to that of the unshocked ejecta $\Gamma_0$, and the time it takes the reverse 
shock to cross the ejecta shell is close to the time it takes the forward-shock to sweep-up a mass 
of ambient medium equal to a fraction $\Gamma_0^{-1}$ of the ejecta mass.

 Thus, the deceleration radius is defined by $M (r_d) = E/\Gamma^2_0$, from where 
\begin{equation}
 r_d \propto  \left( \frac{E}{k \Gamma_0^2} \right)^{1/(3-s)}
\label{rd}
\end{equation}
for an external medium of proton density $n(r) = kr^{-s}$ and the observer-frame deceleration timescale is 
\begin{equation}
 t_d = \frac{r_d}{2c\Gamma^2_0} \;. 
\label{td}
\end{equation}
The condition for a semi-relativistic shock is simply $\Delta < ct_d$, which can be better expressed
as an upper limit on $\Gamma_0$.

 The flux at $t_d$ can be calculated starting from equations (\ref{nui}) -- (\ref{Fo}), with two
modifications. First, the Doppler factor is just the Lorentz factor of the shocked fluid at $t_d$,
which is $\sim \Gamma_0$. Second, $F_o = \Gamma_0 F'_o$ because, owing to the relativistic beaming,
the observer receives emission from a region of opening $\theta = \Gamma_0^{-1}$ that contains
a fraction $\Gamma_0^{-2}$ of the number of electrons $N_e \propto r^3 n$ for a spherical source. 
Therefore
\begin{equation}
 F_o = \Gamma_0 F'_o \propto N_e B \Gamma_0 \propto M \Gamma_0^2 n^{1/2} \propto \Gamma_0^2 n^{3/2} r^3 
\label{fo}
\end{equation}
\begin{equation}
 \nu_i = \Gamma_0 \nu'_i \propto \Gamma_0^4 n^{1/2} \;, \quad \nu_c = \Gamma_0 \nu'_c \propto n^{-3/2} r^{-2}
\label{nuic}
\end{equation}
 
\subsubsection{Homogeneous medium} 

 For $n (r) = n_0$, equations (\ref{rd}) and (\ref{td}) lead to 
\begin{equation} 
 r_p \propto \left(\frac{E}{n_0 \Gamma_0^2}\right)^{1/3} \;,\; 
 t_p \propto \left(\frac{E}{n_0 \Gamma_0^8}\right)^{1/3} \;.
\label{tps0}
\end{equation}
By substituting $r=r_d$ with $r_d$ from equation (\ref{rd}) in equations (\ref{fo}) and (\ref{nuic}),
the spectral characteristics at the peak epoch $t_p = t_d$ satisfy
\begin{equation} 
 F_o (t_p) \propto E n_0^{1/2} ,\;
 \nu_i (t_p) \propto \Gamma_0^4 n_0^{1/2} ,\;
 \nu_c (t_p) \propto E^{-2/3} \Gamma_0^{4/3} n_0^{-5/6} \;.
\end{equation}
Then, for the three orderings of $\nu_i$, $\nu$, and $\nu_c$ that yield a peak for the afterglow light-curve
at the onset of deceleration (i.e. a rising light-curve before $t_d$ and a falling-off flux after that),
we get the peak flux 
\begin{equation}
 F_p (\nu_i < \nu < \nu_c) \propto F_o \nu_i^{\beta}|_{t_p} \propto E \Gamma_0^{4\beta} n^{(\beta+1)/2} 
\label{c1}
\end{equation}
\begin{equation}
 F_p (\nu_i, \nu_c < \nu) \propto F_o \nu_c^{1/2} \nu_i^{\beta-0.5}|_{t_p} \propto 
              E^{2/3} \Gamma_0^{4(\beta-1/3)} n^{(3\beta-1)/6}  
\label{c2}
\end{equation}
\begin{equation}
 F_p (\nu_c < \nu < \nu_i) \propto F_o \nu_c^{1/2}|_{t_p} \propto E^{2/3} \Gamma_0^{2/3} n^{1/12}  \;.
\label{c3}
\end{equation}

 Comparing with equation (\ref{tps0}), the above results show that a variation of $E$ among afterglows 
induces a $F_p - t_p$ positive correlation (unlike the observed anticorrelation). 
For the case in equation (\ref{c1}), varying $n$ and $\Gamma_0$ produces anticorrelations 
$F_p \propto t_p^{-\gamma}$ with 
 \begin{equation}
   \nu_i < \nu < \nu_c : \quad \gamma_n  = \frac{3}{2} (\beta + 1) \;,\; \gamma_\Gamma  = \frac{3}{2} \beta \;.
\end{equation}
thus the average X-ray spectral slope leads to indices $\gamma_{x,n} = 3.1$ and $\gamma_{x,\Gamma} = 
1.56 \pm 0.03$,
respectively. The latter case is consistent with the slope measured for X-ray peaks, however, in that case,
$\beta_o = \beta_x$ and $\gamma_o = \gamma_x$, thus it cannot account for the measured optical peaks slope.
The measured spread of 3 dex in peak epochs would require that $\Gamma_0$ ranges over about 1 dex 
(from eq \ref{tps0}).

For the case given in equation (\ref{c2}), varying $n$ and $\Gamma_0$ leads to the same anticorrelation slope
\begin{equation}
  \nu_i,\nu_c < \nu : \quad \gamma_n = \gamma_\Gamma = \frac{1}{2} (3\beta -1) 
\end{equation}
thus the measured X-ray spectral slope implies $\gamma_x = 1.1$, which is too small.

For the case in equation (\ref{c3}), the corresponding afterglow spectral slope $\beta = 1/2$ is harder 
than usually measured in the X-ray and the resulting indices $\gamma_n = \gamma_\Gamma = 1/4$ are well below 
those measured.

\subsubsection{Wind-like medium} 

 For $n(r) = kr^{-2}$, the equations above become
\begin{equation} 
 r_p \propto \frac{E}{k \Gamma_0^2} \;,\;  t_p \propto \frac{E}{k \Gamma_0^4}
\end{equation}
\begin{equation} 
 F_o (t_p) \propto \Gamma_0^2 k^{3/2} ,\;
 \nu_i (t_p) \propto E^{-1} \Gamma_0^6 k^{3/2} ,\;
 \nu_c (t_p) \propto E \Gamma_0^{-2} k^{-5/2} 
\end{equation}
\begin{equation}
 F_p (\nu_i, \nu_c < \nu) \propto E^{1-\beta} \Gamma_0^{6\beta-2} k^{(3\beta-1)/2}  
\label{C1}
\end{equation}
\begin{equation}
 F_p (\nu_c < \nu < \nu_i) \propto E^{1/2} \Gamma_0 k^{1/4} \;.  
\label{C2}
\end{equation}

 From equation (\ref{C1}), the expected exponents of the $F_p - t_p$ relation are
\begin{equation}
   \nu_i,\nu_c < \nu : \quad \gamma_E = \beta - 1 \;,\; \gamma_k  = \gamma_\Gamma = \frac{1}{2} (3\beta - 1) 
\end{equation}
thus the measured average X-ray spectral slope implies $\gamma_{x,E} = 0$ and $\gamma_{x,k/\Gamma} = 1.1$, 
respectively,
both being less than measured.
 For the case given in equation (\ref{C2}), $\gamma_k  = \gamma_\Gamma = 1/4$, neither of which is 
compatible with observations.
 For $\nu_i < \nu < \nu_c$ the onset of deceleration does not yield a light-curve peak, the afterglow 
flux decreasing even the reverse-shock energizes the ejecta.\footnote{
A light-curve peak would be produced only by the synchrotron peak frequency $\nu_i$ falling through the
observing band; however that is a different light-curve peak mechanism and implies a hard $F_\nu \propto
\nu^{1/3}$ X-ray spectrum before the peak, which is not seen by XRT for any of the afterglows in Table 
\ref{xtable}.}

\subsection{Forward-shock synchrotron emission for a relativistic reverse shock}

 If the ejecta shell is sufficiently thick, the reverse-shock is relativistic and the shocked fluids
move at a Lorentz factor $\Gamma$ that can be well below the $\Gamma_0$ of the incoming ejecta.
It can be shown (Panaitescu \& Kumar 2004) that the reverse-shock reaches the trailing edge of the 
ejecta shell at a radius 
\begin{equation}
 r_d \propto \left( \frac{E \Delta}{k} \right)^{1/(4-s)}
\end{equation}
when the radiating fluid moves at
\begin{equation}
 \Gamma \propto \left( \frac{E} {k \Delta^{3-s}} \right)^{1/(8-2s)}
\label{Gm}
\end{equation}
corresponding to an observer-frame time $t_d = r_d/(2 c\Gamma^2) \simeq \Delta/c$, i.e. the (new) 
observer-frame deceleration timescale is about the lab-frame geometrical thickness of the ejecta shell.
The spectral characteristics at $r_d$ are those of equation (\ref{fo}) and (\ref{nuic}) with $\Gamma$
of equation (\ref{Gm}) instead of $\Gamma_0$. 
Given that $t_d$ depends only on $\Delta$, to derive the peak flux -- peak
epoch relation requires only the dependence of the spectral characteristics on $\Delta$.

 For a homogeneous medium, 
\begin{equation}
 r_d \propto \Delta^{1/4} \;,\;  \Gamma \propto \Delta^{-3/8}
\end{equation}
\begin{equation}
  F_o (t_p) \propto \Delta^0 \;,\; \nu_i (t_p) \propto \Delta^{-3/2} \;,\; \nu_c (t_p) \propto \Delta^{-1/2}
\end{equation}
and the slopes of the $F_p -t_p$ relation induced by the variation of $\Delta$ among afterglows are
\begin{equation}
 \nu_i < \nu < \nu_c:\; \gamma_\Delta = \frac{3}{2} \beta \;,\; 
 \nu_c < \nu < \nu_i:\; \gamma_\Delta =  1/4
\end{equation}
For the first case, the expected slope is $\gamma_x = 1.56 \pm 0.03$, which is consistent with observations, 
but it implies that $\gamma_o = \gamma_x$, hence it does not provide an explanation for the larger 
exponent measured for optical peaks. For $\nu_i, \nu_c < \nu$, the onset of the deceleration does not 
yield a light-curve peak, as the afterglow flux decreases while the reverse-shock crosses the ejecta.

 The same issue (deceleration onset does not yield a light-curve peak) exists for a wind-like medium,
whatever is the $\beta_x > 0$ spectral regime (as measured by XRT).

\section{Conclusions}

 We have identified 31 optical peaks (Fig \ref{opeaks}) with measured redshifts and we fit them with a 
smoothly broken power-law, to determine their peak times and epochs more accurately, and also to assess 
if a time-shift $t_o$ between the beginning of the afterglow emission and the GRB trigger is required.
We found good evidence that $t_o < 0$ for only one afterglow and $t_o > 0$ for five other, while for 
the rest $t_o \neq 0$ does not provide a statistically significant better fit. Because we fit the afterglow
light-curves with power-laws, finding that the afterglow beginning is different than the GRB trigger 
happens mostly when the afterglow rise (with time measured from burst beginning) is curved in log-log space. 
Consequently, such determinations of $t_o$ rely on the untested assumption that the intrinsic afterglow 
rise is a power-law.\footnote{Fortunately, allowing for a non-zero $t_o$ does not change significantly 
the slopes of the optical and X-ray peak flux -- peak epoch relations}

 The $k$-corrected optical light-curves manifest a strong correlation between the peak flux $F_p$ 
and the peak epoch $t_p$. The best fit to the afterglow peaks in the log-log plane is 
$F_p \propto t_p^{-2.15}$. If the peak epochs are shifted by ($\tau$) the middle of the 90 percent 
confidence level on $t_o$, the peak flux - peak epoch best-fit becomes $F_p \propto (t_p-\tau) ^{-2.0}$. 
The probability for the $F_p-t_p$ correlation to occur by chance decreases by a factor 3 when an 
afterglow time-shift (relative to the GRB trigger) is allowed. 

 We have also identified 14 X-ray afterglows (Fig \ref{xpeaks}) that display a light-curve peak. Although XRT 
has monitored several hundreds of X-ray afterglows so far, the X-ray peaks are very rarely seen because 
they are, most often, overshined by the prompt GRB emission. The $k$-corrected X-ray peaks also display 
a peak flux -- peak epoch correlation: $F_p \propto t_p^{-1.6}$, which is slower than for optical peaks,
and whose index changes little if a shift of the afterglow beginning is allowed. There is a 50 percent 
chance that the distributions of optical and X-ray peak epochs (Fig \ref{oxpeaks}) are the same.

 The correlations identified for optical and X-ray peaks represent the bright and late edge of the entire 
distribution of peaks in the $F_p - t_p$ plane, as peaks that are dimmer or occur earlier are more likely
to be missed.
Perhaps, the only model-independent conclusion that can be drawn from the existence of a bright and late
edge in the distribution of peaks is the existence of an upper limit to the energy that afterglows radiate.
Two thirds of optical and X-ray peaks decay faster than $t^{-1}$ after the peak, thus $F_p t_p$ is a 
good measure of the radiative output for most afterglows. Then, the peak correlation having a slope
larger than unity (1.6 in X-rays, 2.0 in the optical) implies that later peaks radiate less energy than 
earlier ones.

 To reach more detailed conclusions, we have investigated two likely models for light-curve peaks, 
both pertaining to the synchrotron emission from the forward-shock the energizes the burst ambient medium. 
The slope of the bright-edge of the $F_p-t_p$ distribution enables a test of each model if it is assumed 
that that anticorrelation is induced by the variation of one parameter among the set of afterglows with peaks.

 One model is that of an off jet-aperture observer location, the afterglow peak occurring when the jet 
has decelerated enough that its emission is relativistically beamed toward the observer. In this
model, jets seen at a larger angle peak later and dimmer. We have derived the $F_p - t_p$ correlation 
induced by a variation of the observer offset angle, for both conical and laterally-spreading jets, 
and for a homogeneous or a wind-like medium, and we have found that a conical jet (and either type of 
medium) may explain the observed X-ray peak correlation, given the measured average X-ray spectral 
slope $\overline{\beta_x} = 1.0$. If the optical peaks correlation is, indeed, steeper than for the 
X-ray peaks, that would require a wind-like ambient medium.  
 
 The other model for afterglow peaks considered here is the onset of the forward-shock deceleration,
occurring when the reverse shock has crossed the ejecta shell, the afterglow peak being due to a change
in the dynamics of the forward-shock, caused by a decrease in the rate at which energy is transferred 
from the ejecta to the forward-shock. We have derived the $F_p - t_p$ correlation expected for a 
semi-relativistic reverse-shock ($F_p$ and $t_p$ depend on the ejecta initial Lorentz factor and
the ambient medium density) and for a relativistic reverse-shock ($F_p$ and $t_p$ depend only on the 
geometrical thickness of the ejecta shell) and have found that this model can explain the $F_p - t_p$ 
relation measured for X-ray peaks if the ambient medium is homogeneous and if X-ray is below the
synchrotron cooling frequency (for either type of reverse-shock), but it underpredicts the slope
of the optical peak flux -- peak epoch relation. 

  There are three important caveats for any conclusion drawn from the slope of the optical and X-ray
afterglow peaks. 

 The first caveat is that the sample of X-ray peaks is small, hence the slope $\gamma_x = 1.6$ 
of the $F_p - t_p$ relation is still quite uncertain. 
The X-ray afterglow 060614 displays a plateau around the epoch of the optical peak and, thus, was not 
included in the sample of X-ray peaks. However, if an X-ray peak is assigned at the epoch of the optical 
peak, then the best-fit to the peak fluxes and epochs of the resulting set of 15 X-ray afterglows would 
have a slope $\gamma_x = 2.1$, compatible with that measured in the optical ($\gamma_o = 2.0$).

 The second caveat is that requiring a model to account for the slopes of both optical and X-ray peaks 
is justified only if the afterglow emissions at these two frequencies arise from the same population of 
electrons.
A simple test for the common origin of the optical and X-ray afterglow emission is the simultaneity 
of the light-curve peaks. Most of optical peaks occurred while the X-ray afterglow was dominated by 
the prompt emission, hence the achromaticity of optical peaks is, in general, impossible to establish. 
The second optical peak of 100901 appears also in the X-ray, the optical and X-ray peaks of 100418 
seem achromatic, but those of 110213 are chromatic. There are four optical peaks (060206, 060614, 
070802, 081029) for which the X-ray data suggest the existence of a simultaneous peak, although they 
do not show it clearly, and one (110205) achromatic optical peak. For two X-ray peaks (080310, 080319C), 
the optical data allow for a simultaneous optical peak occurring at the end of a light-curve plateau. 
All in all, there is evidence that optical and X-ray light-curve peaks occur simultaneously more 
often than not, i.e. in favour of a common origin of the optical and X-ray afterglow emissions.

 The third caveat is that we have tested only the ability of one model parameter to induce the observed
$F_p - t_p$ anticorrelation but, in each model, there are other parameters that could alter the slope
of that correlation. For the off-aperture observer model, where the fundamental parameter that drives
the peak correlation is the observer's offset angle, the jet energy per solid angle and the ambient 
medium density may be seen as secondary parameters whose variation among afterglows only introduces
some scatter around the peak correlation induced by the main parameter.
However, for the deceleration-onset model, there are two basic parameters that can induce an  $F_p - t_p$ 
anticorrelation (the ambient density and the ejecta initial Lorentz factor, or the ejecta shell thickness), 
the resulting slope of the $F_p - t_p$ relation depending on the (unknown) width of the distribution of 
those two parameters.

 Moreover, given that both models for afterglow peaks considered here yield $F_p -t_p$ relation slopes 
that are either smaller or larger than measured for X-ray and optical peaks, it is possible that a
peak relation of intermediate slope results from combining both peak mechanisms, some peaks being
due to an off-jet observer location while others are caused by the onset of deceleration. That would 
make it even harder to use the $F_p - t_p$ relations identified here for model testing.

\vspace*{-5mm}
\section*{Acknowledgments}
 This work was supported by an award from the Laboratory Directed Research and Development program
 at the Los Alamos National Laboratory and made use of data supplied by the UK Science Data Center
 at the University of Leicester.

\end{document}